\documentclass[12pt]{article}

\usepackage{graphics}
\usepackage{epsfig}

\newcommand{\nn}{\nonumber}

\newcommand{\bq}{\begin{eqnarray} }
\newcommand{\eq}{\end{eqnarray} }
\newcommand{\e}{\epsilon}

\begin{document}\begin{titlepage}

\begin{flushright}
OSU-HEP-04-5\\
May 2004\\
\end{flushright}

\vspace*{1.5cm}
\begin{center}
{\Large {\bf Fermion Mass Hierarchy and \\Electric Dipole Moments\\[-0.05in]
} }

\vspace*{2cm}
 {\large {\bf K.S. Babu\footnote{E-mail address:
 babu@okstate.edu} and
 Ts. Enkhbat\footnote{E-mail address: enkhbat@okstate.edu}
 }}

 \vspace*{1cm}
{\it Department of Physics, Oklahoma State University\\
Stillwater, OK~74078, USA }
\end{center}

 \vspace*{1.5cm}

\begin{abstract}

We show that in supersymmetric models with a gauged flavor
symmetry explaining the hierarchy of fermion masses and mixings,
the electron, muon, neutron and the deuteron acquire sizable
electric dipole moments (EDM) through loop diagrams involving the
flavor gaugino/gauge boson near the Planck scale. These EDMs are
proportional to the phases in the fermion Yukawa couplings and are
typically much larger than the neutrino seesaw induced EDM for the
leptons. In a popular class of models based on anomalous $U(1)$
flavor symmetry of string origin, we find $d_e\sim
(10^{-26}-10^{-27})~ e$ cm, $d_\mu\sim (10^{-24}-10^{-26})~e$ cm,
$d_n\sim10^{-27}~e$ cm, and $d_D \sim (10^{-26}-10^{-27})~e$ cm,
which are within reach of next generation experiments.
\end{abstract}

\end{titlepage}
\newpage

\section{Introduction}

The observed hierarchy in the masses and mixings of quarks and
leptons is one of the most puzzling features of Nature. A
plausible explanation is provided by flavor--dependent gauge
symmetries. In a popular class of such models one extends the
Standard Model (SM) to include a family--dependent $U(1)$ factor.
Upon spontaneous breaking of the $U(1)$ symmetry effective Yukawa
couplings of the form $y_{ij}\e^{n_{ij}}f_if^c_jH$ are induced,
where $\e\sim 0.2$ is a small parameter, and $n_{ij}$ are positive
integers related to the family dependent $U(1)$ charges. Even when
the fundamental Yukawa couplings $y_{ij}$ are all of order one, a
hierarchical spectrum is realized due to the suppression in powers
of $\e$ \cite{FrNl}. Such flavor $U(1)$ symmetries can be
naturally identified with the anomalous $U(1)_A$ symmetry of
string theory \cite{GS}. Models using anomalous $U(1)_A$ symmetry
for fermion mass and mixing hierarchy abound in the literature
\cite{IbanezU1}--\cite{beg}. Most models of this type also assume
low energy supersymmetry (SUSY) to stabilize the Higgs boson mass.
Novel phenomena which are amenable to experimental tests can arise
in such contexts. The purpose of this paper is to analyze one such
effect, viz., the electric dipole moments of elementary fermions
\cite{physicstoday}.

Low energy supersymmetry can potentially induce excessive flavor
violation in processes such as $K^0-\overline{K^0}$ mixing and
$\mu\rightarrow e\gamma$ decay if the soft supersymmetry breaking
Lagrangian takes its most general form. This potential problem is
usually avoided by assuming a universal form for the soft SUSY
breaking terms. Even with universality, the $CP$--violating phases
present in the soft SUSY breaking Lagrangian can induce electric
dipole moments (EDM) for the neutron and the electron at a level
exceeding the current experimental limits. These effects have been
extensively studied in the literature
\cite{Ellis1}--\cite{Ellis3}. We will assume in the present work a
universal SUSY breaking spectrum that is also $CP$--invariant so
that excessive EDMs are not induced from the fundamental soft SUSY
breaking parameters.

The EDMs that we find in the context of models of fermion mass
hierarchy are induced purely by complex Yukawa couplings. The
phases in the Yukawa couplings are believed to be the source for
the observed $CP$--violation in the $K$ and $B$ meson systems (CKM
$CP$--violation). It is thus reasonable to assume all Yukawa
couplings, including the leptonic Yukawa couplings, to be complex.
As we will see, it is natural that the flavor $U(1)$ gauge
symmetry responsible for explaining the fermion mass hierarchy
breaks spontaneously at a scale $M_F$ slightly below the
fundamental Plank (or string) scale, $M_F\sim M_{st}/50$. In the
momentum regime $M_F<\mu<M_{st}$ the flavor gauge sector will be
active and will contribute to flavor violation in the squark and
slepton sectors. These effects would survive down to the SUSY
breaking scale and can lead to observable phenomena. In a previous
paper we have studied leptonic rare decays $\mu\rightarrow
e\gamma$ and $\tau\rightarrow \mu\gamma$ induced by the flavor
gauge sector \cite{beg}. With complex Yukawa couplings, this
flavor violation will also lead to EDMs for the electron ($d_e$),
muon ($d_\mu$), the neutron ($d_n$), and the deuteron ($d_D$) even
with universal and $CP$--conserving soft SUSY breaking terms at
the string scale. In a popular class of anomalous $U(1)$ models
which explains the fermion mass hierarchy, including bi--large
neutrino mixing, we find $d_e\sim (10^{-26}-10^{-27})~e$ cm,
$d_\mu\sim (10^{-24}-10^{-26})~e$ cm, $d_n\sim 10^{-27}~e$ cm, and
$d_D \sim (10^{-26}-10^{-27})~e$ cm, which are within reach of
next generation experiments. There are proposals to improve the
current limit on electron EDM, $|d_e|\leq 1.6\times 10^{-27}~e$ cm
\cite{Regan}, by about two to four orders of magnitude
\cite{DeMille, lamoreaux}. It is expected that the current limit
on the muon EDM, $|d_\mu|\leq 1.9\times 10^{-18}~e$ cm, will be
improved by six orders of magnitude or even more in the not too
distant future \cite{MuonEDM}. There are also proposals which
would improve the current neutron EDM limit from $|d_n|\leq
6.3\times 10^{-26}~e$ cm \cite{Harris} by a factor of 5
\cite{NeutronEDM}. The deuteron EDM is expected to be probed to
the level of $10^{-27}~e$ cm in the near future \cite{deuteron}.
Supersymmetry may reveal itself in these experiments before direct
discovery at LHC, if the current ideas of solving the fermion mass
hierarchy problem are correct.

Lepton EDMs may arise even without flavor gauge symmetry from
complex neutrino Yukawa couplings responsible for the seesaw
mechanism in the context of low energy SUSY. This effect has
received much attention recently \cite{Ellis3, Masina, peskin}. We
have computed such effects for $d_e$ and $d_\mu$, but found them
to be much less significant compared to the flavor $U(1)$ induced
effects. For example, we find $d_e\sim 10^{-29}~e$ cm for large
$\tan\beta$ from the neutrino Yukawa coupling effects, to be
compared with $d_e\sim10^{-26}~e$ cm from the flavor $U(1)$
sector.  Similar effects from GUT threshold has been studied in
Ref. \cite{Borzumati}.

The paper is organized as follows. In Section 2 we review the
class of models based on anomalous $U(1)_A$ symmetry for fermion
mass hierarchy \cite{beg}. In Section 3 the EDMs induced by the
flavor $U(1)_A$ gauge sector is analyzed. In 3.1 a qualitative
discussion of the radiative corrections to the soft masses and
$A$--terms is given. In 3.2 we present our full numerical results
for the EDMs. Section 4 has our conclusions. In Appendix \ref{a1}
we give the relevant expressions for the $\beta$--functions for
the soft SUSY breaking parameters including corrections from the
$U(1)_A$ gauge sector. In Appendix \ref{a2} the fermion mass fit
for the model used in the numerical analysis is presented.
Appendix \ref{a3} lists the formulas needed for the calculation of
EDMs.

\section{Fermion Masses and Anomalous $U(1)$ Symmetry}

In this section we review briefly the idea of explaining fermion
mass hierarchy with a flavor dependent $U(1)$ symmetry. We focus
on a specific class of anomalous $U(1)_A$ models discussed in Ref.
\cite{beg} to address the fermion EDM.  Most models of Ref.
\cite{IbanezU1}--\cite{Kobayashi} will also fall into this
category and will lead to similar results.  In these models
families are distinguished by their anomalous $U(1)$ charges. The
$U(1)_A$ symmetry is broken spontaneously by an MSSM singlet
flavon field $S$ which acquires a vacuum expectation value (VEV)
slightly below the string scale $M_{st}$. This provides a small
expansion parameter $\e=\langle S\rangle/M_{st}$ needed for
explaining the fermion mass hierarchy. $U(1)$ invariance forbids
renormalizable Yukawa couplings for the light families, but would
allow them through effective nonrenormalizable couplings
suppressed by a factor $(S/M_{st})^{n_{ij}}$ (for the fermion mass
operator connecting flavors $i$ and $j$) with $n_{ij}$ being
positive integers. Even with all couplings being of order one,
hierarchical masses for different flavors are naturally realized
\cite{FrNl}. Although this mechanism will work with any flavor
$U(1)$, anomalous $U(1)$ models are attractive since they would
also provide a natural understanding for the smallness of
$\e\sim0.2$ \cite{IbanezU1}, which arises from the one--loop
induced Fayet--Illiopoulos $D$--term \cite{DSW}.

Consider the following fermion mass matrices studied in Ref
\cite{beg}:

\begin{eqnarray}\label{massM1}
&&M_u\sim \langle
H_u\rangle\pmatrix{\epsilon^{\,8}&\epsilon^{\,6}&\epsilon^{\,4}\cr
\epsilon^{\,6}&\epsilon^4&\epsilon^2\cr\epsilon^{\,4}&\epsilon^2&1}\,,\hspace{1.cm}
M_d\sim \langle H_d\rangle\epsilon^p
\pmatrix{\epsilon^{\,5}&\epsilon^{\,4}&\epsilon^{\,4}\cr
\epsilon^3 &\epsilon^2&\epsilon^2\cr\epsilon&1&1},\nn\\
\nn\\
\vspace{.5cm} &&M_e\sim \langle
H_d\rangle\epsilon^p\pmatrix{\epsilon^{\,5}&\epsilon^3&\epsilon\cr
\epsilon^{\,4}&\epsilon^2&1\cr\epsilon^{\,4}&\epsilon^2&1}\,,\hspace{1.cm}
M_{\nu_D}\sim \langle H_u\rangle\epsilon^s
\pmatrix{\epsilon^2&\epsilon &\epsilon \cr \epsilon
&1&1\cr\epsilon&1&1},\nn\\
\nn\\
 \vspace{.5cm} &&M_{\nu^c}\sim M_R
\pmatrix{\epsilon^2&\epsilon&\epsilon\cr
\epsilon&1&1\cr\epsilon&1&1}\,\hspace{.5cm}\Rightarrow
\hspace{.5cm} M^{light}_\nu\sim \frac{{\langle
H_u\rangle}^2}{M_R}\epsilon^{2s}
\pmatrix{\epsilon^2&\epsilon&\epsilon\cr
\epsilon&1&1\cr\epsilon&1&1}.
\end{eqnarray}
Here $M_u$, $M_d$ and $M_e$ are the up--quark, the down--quark and
the charged lepton mass matrices in the basis $fM_ff^c\,(f=u,
\,d,\,e,\,\nu)$. Complex order one coefficients multiplying each
entry of the matrices are not shown. $M_{\nu_D}$ and $M_{\nu^c}$
are the neutrino Dirac and Majorana mass matrices. The light
neutrino mass matrix $M_{\nu}^{light}$ is derived from the seesaw
mechanism. $p$ and $s$ are integers and are chosen differently for
different values of $\tan\beta=\left\langle
H_u\right\rangle/\left\langle H_d\right\rangle$. The choice of
$p=0,\,1,\,2$ corresponds to large ($\sim50$), medium ($\sim20$)
and small ($\sim5$) values of $\tan\beta$ respectively. The quark
and lepton masses and mixings arising from Eq. (\ref{massM1}) are
fully consistent with experimental observations if $\e\sim0.2$.
Note that the CKM mixing angles are small, while the leptonic
mixing angles relevant for solar and atmospheric neutrino
oscillations are of order unity. The lopsided nature of the
matrices $M_d$ and $M_e$ of Eq. (\ref{massM1}) enables such a
disparity to be realized \cite{lopsided}.

The superpotential which would lead to the Yukawa coupling
structure in Eq. (\ref{massM1}) has the following general form
near the fundamental scale $M_{st}$:
\begin{eqnarray}\label{superP1}
W&=&\sum_{f}\frac{y_{ij}^f}{n^f_{ij}!}  f_iHf^c_j
\left(\frac{S}{M_{st}}\right)^{n^f_{ij}}+{{M_{R}}_{ij} \over 2~
n^{\nu^c}_{ij}!} \nu^c_i
\nu^c_j\left(\frac{S}{M_{st}}\right)^{n^{\nu^c}_{ij}}+\mu H_u
 H_d+W_{A}\left(S, X_k\right)\, .
\end{eqnarray}
Here $i,j=1,2,3$ are the generation indices and $n^f_{ij}\,(f=u,
\,d,\,e,\,\nu)$ are positive integers fixed by the choice of the
$U(1)_A$ charge assignment. We choose all charges to be integers
with the charge of $S$ being negative. $y^f_{ij}$ are the Yukawa
couplings which we take to be complex and of order one for all
$i,\,j$. $H$ stands for the MSSM Higgs doublets $H_u$ and $H_d$.
$S$ is the MSSM singlet flavon field whose VEV determines the
expansion parameter $\e$. $W_A$ in Eq. (\ref{superP1}) contains
MSSM singlet fields $X_k$ which would be needed  for anomaly
cancelation. The $U(1)$ charge assignment shown in Table
\ref{tcharge} will lead to the texture of Eq. (\ref{massM1}).
\begin{table}[t]
\begin{center}
\begin{tabular}{|c|c|c|}\hline
\rule[5mm]{0mm}{0pt}Field& $U(1)_A$ Charge& Charge
notation\\\hline
\rule[5mm]{0mm}{0pt}$Q_1$, $Q_2$, $Q_3$& \,\,$ 4, \,2, \,0$ &$q^Q_i$\\
\rule[5mm]{0mm}{0pt}$L_1$, $L_2$, $L_3$&\,\,$ 1+s,\, s,
\,s$&$q^L_i$\\
\rule[5mm]{0mm}{0pt}$u^c_1$, $u^c_2$, $u^c_3$&$ 4,\, 2,\, 0$&$q^u_i$\\
\rule[5mm]{0mm}{0pt}$d^c_1$, $d^c_2$, $d^c_3$&$1+p,\, p,\,
p$& $q^d_i$\\
\rule[5mm]{0mm}{0pt}$e^c_1$,$e^c_2$,$e^c_3$&\,\,$ 4+p-s,\, 2+p-s,\, p-s$\,\,&$q^e_i$\\
\rule[5mm]{0mm}{0pt}$\nu^c_1$, $\nu^c_2$,
$\nu^c_3$ &$ 1,\, 0,\, 0$&$q^\nu_i$\\
\rule[5mm]{0mm}{0pt}$H_u$, $H_d$, $S$&$ 0,\, 0,\, -1$&$(h, \bar{h}, q_s)$\\
\hline
\end{tabular}
\caption{\footnotesize The flavor $U(1)_A$ charge assignment for
the MSSM fields and the flavon field $S$ in the normalization
where $q_s=-1$. In the third column we list the notation for the
charges used in the paper.}
  \label{tcharge}
\end{center}
\end{table}

The $U(1)_A$ anomalies are cancelled by the Green--Schwarz
mechanism \cite{GS} which requires
\begin{eqnarray}\label{AnomalyGS}
\frac{A_1}{k_1}=\frac{A_2}{k_2}=\frac{A_3}{k_3}=\frac{A_{F}}{3k_F}=\frac{A_{gravity}}{24}\,.
\end{eqnarray}
Here $A_1$, $A_2$, $A_3$, $A_F$ and $A_{gravity}$ are
$U(1)_Y^2\times U(1)_A$, $SU(2)_L^2\times U(1)_A$,
$SU(3)_C^2\times U(1)_A$, $U(1)^3_A$ and $(Gravity)^2\times
U(1)_A$ anomaly coefficients. All other anomalies (such as
$U(1)_A^2\times U(1)_Y$) must vanish. $k_i\,(i=1,2,3)$, $k_F$ are
the Kac-Moody levels, with the Non--Abelian levels $k_2$ and $k_3$
being integers. The factor $1/3$ in front of the cubic anomaly
$A_F$ has a combinatorial origin owing to the three identical
$U(1)_A$ gauge boson legs.

We require string unification of all the gauge couplings including
that of the $U(1)_A$, $g_F$, at the fundamental scale $M_{st}$
\cite{Ginsparg}:
\begin{eqnarray}\label{UnifGS}
k_ig_i^2=k_Fg_F^2=2g_{st}^2.
\end{eqnarray}
For a clear discussion of the coefficients in Eqs.
(\ref{AnomalyGS})--(\ref{UnifGS}) see Ref. \cite{cvetic}.  Here
$g_i$ are the $U(1)_Y$, $SU(2)_L$ and $SU(3)_C$ gauge couplings
for $i=1,\,2,\,3$. With the choice $k_2=k_3=1$, consistency with
the observed unification of gauge couplings within MSSM would
require $k_1=5/3$, corresponding to the $SU(5)$ normalization of
hypercharge. This will be an automatic consequence of the
Green--Schwarz anomaly cancelation conditions with our choice of
texture. The small discrepancy (in the absence of a covering GUT
group) between the unification scale derived from low energy data
and the string scale may be understood in the context of
$M$--theory by making use of the radius of the eleventh dimension
\cite{witten}.

With $k_2=k_3=1$ we find from Table \ref{tcharge}, $A_2=(19+3s)/2$
and $A_3=(19+3p)/2$. Eq. (\ref{AnomalyGS}) then requires $p=s$,
i.e., a common exponent for the charged lepton and the neutrino
Dirac Yukawa coupling matrices. With $p=s$, the condition
$A_1/k_1=A_2/k_2$ fixes $k_1$ to be $5/3$, consistent with $SU(5)$
unification. Note also that the charges given in Table
\ref{tcharge} become compatible with $SU(5)$ unification. Since
${\rm Tr}(Y) = 0$ for the fermion multiplets of $SU(5)$, and since
the Higgs doublets carry zero $U(1)_A$ charge, the anomaly
coefficient $[U(1)_A]^2 \times U(1)_Y$ vanishes, as required. The
last equality in Eq. (\ref{AnomalyGS}) requires
\begin{eqnarray}\label{Agravity}
A_{gravity}=\mbox{Tr}\left(q\right)=12(19+3p).
\end{eqnarray}
This cannot be satisfied with the MSSM fields alone, since
$\mbox{Tr}(q)_{MSSM}=5(13+3p)$, which does not match Eq.
(\ref{Agravity}). We cancel this anomaly by introducing MSSM
singlet fields $X_k$ obeying
$\mbox{Tr}\left(q\right)_X=A_{gravity}-\mbox{Tr}\left(q\right)_{MSSM}=163+21p$.
If all the $X_k$ fields have the same charge equal to $+1$, they
will acquire masses of order $M_{st}\e^2$ through the coupling
$X_kX_kS^2/M_{st}$ and will decouple from low energy theory. For
other choices of the charge of $X_k$ these masses can be
different. For example, if the charge is equal to $+1/2$, their
masses will be of order $M_{st}\e$; if the charge is $+2$ the
masses will be of order $M_{st}\e^4$. We will consider only the
case where the $X_k$ fields have charge $+1$.

With the charges of all fields fixed, we are now in a position to
determine the $U(1)_A$ charge normalization so that
$g_F^2=g_2^2=g_3^2$ at the string scale, (We take $k_2=k_3=1$.)
This normalization factor, which we denote as $|q_s|$, is given by
$|q_s|=1/\sqrt{k_F}$. All the charges given in Table \ref{tcharge}
are to be multiplied by $|q_s|$. From the Green--Schwarz anomaly
cancelation condition $A_F/(3k_F)=A_2/k_2$, we have
\begin{eqnarray}\label{Ax}
\frac{\mbox{Tr}\left(q^3\right)}{3k_F}=\frac{19+3p}{2k_2},
\end{eqnarray}
from which we find the normalization of the $U(1)_A$ charge
$|q_s|=1/\sqrt{k_F}$ to be
\begin{eqnarray}\label{anomalyGS}
|q_s|=\left(0.179,\,0.186,\,0.181\right)\,\, \mbox{for $p=(0, 1,
2$)}\, .
\end{eqnarray}

The Fayet--Iliopoulos term for the anomalous $U(1)_A$, generated
through the gravitational anomaly, is given by \cite{DSW}
\begin{eqnarray}
\xi=\frac{g_{st}^2M_{st}^2}{192\pi^2}|q_s|A_{gravity}\,,
\end{eqnarray}
where $g_{st}$ is the unified gauge coupling at the string scale
(see Eq. (\ref{UnifGS})). By minimizing the potential from the
$U(1)_A$ $D$--term
\begin{eqnarray}\label{Dterm}
V=\frac{|q_s|^2g_F^2}{8}\left(\frac{\xi}{|q_s|}-|S|^2+\sum_aq_a^f|\tilde{f}_a|^2+\sum_k
q_k^X|X_k|^2\right)^2,
\end{eqnarray}
in such a way that supersymmetry remains unbroken, one finds for
the VEV of $S$
\begin{eqnarray}\label{Eps}\epsilon=\langle
S\rangle/M_{st}=\sqrt{g_{st}^2A_{gravity}/192\pi^2}.\end{eqnarray}
For the fermion mass texture in Eq. (\ref{massM1}), corresponding
to the $U(1)_A$ charges given in Table \ref{tcharge}, we find
\begin{eqnarray}\label{epsilon}
\epsilon=\left(0.177,\,0.191,\,0.204\right)\,\, \mbox{for $p=(0,
1, 2$)}\, .
\end{eqnarray}
The masses of the $U(1)_A$ gauge boson and the corresponding
gaugino are obtained from $M_F=|q_s|g_F\langle S\rangle/\sqrt{2}$
and found to be
\begin{eqnarray}\label{MAM1}
M_F=\left(\frac{M_{st}}{54.5},\,
\frac{M_{st}}{52.5},\,\frac{M_{st}}{53.9}\right)\, \mbox{for
$p=(0, 1, 2$)}\, .
\end{eqnarray}
In the momentum range below $M_{st}$ and above $M_F$, these gauge
particles will be active and will induce flavor dependent
corrections to the sfermion soft masses and the $A$--terms. It is
these effects which induce EDMs for the electron, muon and the
neutron at low energies.

\section{Electric Dipole Moments from Anomalous $U(1)$}

In the Standard Model the electric dipole moments of the electron,
muon and the neutron are predicted to be extremely small and
beyond reach of planned experiments. In the presence of low energy
supersymmetry these EDMs can exceed the current experimental
limits if soft SUSY breaking parameters are complex
\cite{Ellis1}--\cite{Ellis3}. To focus on the anomalous $U(1)$
induced effects we shall adopt the minimal supergravity scenario
with universal and $CP$--conserving soft SUSY breaking parameters.
Specifically, at the string scale we assume a universal scalar
mass $m_0$, a common gaugino mass $M_{1/2}$, and trilinear
$A$--terms proportional to their respective Yukawa couplings. We
assume $m_0$, $M_{1/2}$ and $A_0$, and the Higgs mass parameters
$\mu$ and $B\mu$ to be real. Thus the only source of
$CP$--violation is in the complex Yukawa couplings. This is needed
for the CKM $CP$--violation in the quark sector and it is natural
to assume that the leptonic Yukawa couplings are complex as well.
In Sec. 3.1 we give a qualitative estimate of the EDMs induced by
the radiative corrections involving the $U(1)_A$ gauge sector in
such a SUSY context. Our numerical results are presented in Sec.
3.2.

\subsection{$U(1)_A$ Correction to the Soft Parameters and EDM}

We now give approximate expressions for the $U(1)_A$ gauge sector
RGE corrections to the soft parameters between the string scale
and the $U(1)_A$ breaking scale $M_F$. The full RGE expressions
for the soft parameters in the presence of higher dimensional
operators as in Eq. (\ref{superP1}) have been derived in Ref.
\cite{beg}. In Appendix \ref{a1} we summarize the relevant
expressions. The $U(1)_A$ corrections to the soft masses for the
left--handed slepton are obtained from Eq. (\ref{betamf}) to be
\begin{eqnarray}\label{delMs}
&&\delta\left(m_{\tilde{L}}^2\right)^{A}_{ij}\simeq\left(4(q^L_iM_{\lambda_F})^2
-q^L_i m_0^2\mbox{Tr}\left(q\right)\right)(|q_s|g_F)^2\delta_{ij}
\frac{{\rm log}\left(M_{st}/M_F\right)}{8\pi^2},
\end{eqnarray}
and a similar expression for the right--handed slepton masses with
the interchange $(\tilde{L},\,q^L)$ \newline $\rightarrow
(\tilde{e},\,q^e)$. There are analogous corrections in the squark
sector. The corrections to the $A$--terms are obtained from Eq.
(\ref{beta2}) as
\begin{eqnarray}\label{delA} \delta A^e_{ij}\simeq-M_{\lambda_F}{g_F}^2
Y^e_{ij} Z^e_{ij}\frac{{\rm
log}\left(M_{st}/M_F\right)}{4\pi^2}\,,
 \end{eqnarray}
where $Z_{ij}$ are biliear combinations of the flavor charges
given by \cite{beg}
\begin{eqnarray}\label{Zfactor}
Z^e_{ij}&=&q^L_i q^e_j+q^L_i {\bar h} +q^e_j{\bar
h}+n^e_{ij}q_s(q^L_i+q^e_j+ {\bar
h})+\frac{1}{2}n^e_{ij}(n^e_{ij}-1)q_s^2.
\end{eqnarray}
Numerical values of $Z^e_{ij}$ for our model are given in Eq.
(\ref{Ze1}) of Appendix \ref{a1}. Note that these corrections,
Eqs. (\ref{delMs}) and (\ref{delA}), are flavor dependent. Due to
the flavor dependent nature of these corrections, the fermion and
the corresponding sfermion mass matrices cannot be diagonalized
simultaneously. This was the source of the flavor violation
studied in Ref. \cite{beg}. For the same reason, with complex
Yukawa couplings $Y^f_{ij}$, nonzero EDMs for the fermions will be
induced.

Let us now estimate the EDM of the electron arising from the
corrections in Eqs. (\ref{delMs}) and (\ref{delA}). There are
three flavor dependent matrices in the leptonic sector, not
including the neutrino Yukawa matrix $Y^\nu$. They are the
leptonic Yukawa matrix $Y^e$ and the matrices of $U(1)_A$ charges
$\hat{Q}^L=diag\left(1+p,\,p,\,p\right)$ and
$\hat{Q}^e=diag\left(4,\,2,\,0\right)$ for the lepton doublets and
singlets (see Table \ref{tcharge}). In the mass eigenbasis for the
charged leptons $\hat{Q}^L$ and $\hat{Q}^e$ will develop complex
off diagonal entries, with the phases arising from $Y^e$ through
the unitary matrices that diagonalize $Y^e$. This is the basic
source for the EDM. The corrections given in Eq. (\ref{delMs})
will generate EDM of the electron through the product of slepton
mixings in $(1i)_{LL}$, $(ii)_{LR}$ and $(i1)_{RR}$ (for $i=2,3$).
The induced EDM will be $d_e\propto\mbox{Im}\left[\left(U^\dagger
\hat{Q}^LY^e\hat{Q}^eV^\dagger\right)_{11}\right]$, where $U$ and
$V$ are unitary matrices which diagonalize $Y^e$,
$Y^e=UY^e_{daig}V^\dagger$. There are additional corrections which
are quadratic in $\hat{Q}^L$ and $\hat{Q}^e$. The corrections to
the $A$--terms in Eq. (\ref{delA}) will also induce EDM directly
through $(LR)$ mixings. Combining these effects with the formula
for the EDM given in Eq. (\ref{FormulEDM}) of Appendix \ref{a3} we
arrive at the following approximate expression for $d_e$:
\begin{eqnarray}\label{edmapprx1}
d_e/e\simeq\frac{\alpha v_dM_{\tilde{B}}}{8\pi \cos^2\theta_W}
\,\frac{1}{m_{\tilde{l}}^2}
A\left(\frac{M^2_{\tilde{B}}}{m_{\tilde{l}}^2}\right)\frac{(|q_s|g_F)^2{\rm
log}\left({M_{st}}/{M_F}\right)}{8\pi^2}\sum_{i=2,3}
\left[C^m_i+C^A_i\right]\mbox{Im}\left[\frac{Y^e_{1i}Y^e_{i1}}{Y^e_{ii}}\right],
\end{eqnarray}
where $C^m_i$ and $C^A_i$ denote the contributions from the soft
masses and the $A$--terms respectively. They are given by
\begin{eqnarray}\label{deltal1}
&&C^m_i=\frac{(|q_s|g_F)^2{\rm
log}\left({M_{st}}/{M_F}\right)}{8\pi^2}\frac{m_0^4\left(A_{0}-|\mu|\tan\beta
\right)}{m_{\tilde{l}}^6}H^L_iH^R_i,\nn\\
&&H^L_i=4\left(M_{1/2}/m_0\right)^2\left((q^L_i)^2-(q^L_1)^2\right)-(q^L_i-q^L_1){\rm
Tr}q,\nn\\
&&C^A_i=2\frac{M_{1/2}}{m_{\tilde{l}}^2}\left(Z^e_{i1}-Z^e_{11}\right).
\end{eqnarray}
Here $H^R_i$ is obtained from $H^L_i$ by the replacement
$q^L_i\rightarrow q^e_i$. $m_{\tilde{l}}$ is the average slepton
mass and $M_{\tilde{B}}$ is the Bino mass. The function $A\left(
X\right)$ is given in Eq. (\ref{FuncEDM}) in Appendix \ref{a3}. We
see explicitly that the complex Yukawa couplings along with
nonuniversal $U(1)_A$ charges lead to nonzero EDM.

To estimate the size of this effect we choose the approximations
$m_0=M_{1/2}\simeq M_{SUSY}$. Following the mass matrices given in
Eq. (\ref{massM1}) we take $|Y^e_{ij}|\simeq \e^{n^e_{ij}+p}$. We
consider here only the contribution from the $(13)$ mixing, since
the $U(1)_A$ charge difference is the largest between the first
and the third generations in $\hat{Q}^e$. Then we find
\begin{eqnarray}\label{edmapprx2}
d_e/e&\sim&\left(10^{-27}\mbox{cm}\right)\times
\left(\frac{500\mbox{GeV}}{m_{\tilde{l}}}\right)^2\nn\\&\times&
M_{\tilde{B}}\left(O(10)\frac{M_{SUSY}^4(|\mu|\tan\beta)}{m_{\tilde{l}}^6}+O(1)\frac{M_{SUSY}}{m_{\tilde{l}}^2}\right)\,\mbox{Arg}
\left[Y^e_{13}Y^e_{31}\right].
\end{eqnarray}
From this estimate we see that the electron EDM induced by the
$U(1)_A$ gauge corrections is in the experimentally interesting
range and already puts constraint on the soft SUSY breaking
parameters. The actual numerical result is quite sensitive to the
choice of $m_0$ and $M_{1/2}$. In our numerical calculations we
have chosen $m_0=M_{1/2}/4.4$ for low $\tan\beta$ for cosmological
reason. In this case the $O(10)$ coefficient in Eq.
(\ref{edmapprx2}) will be reduced to an $O(1)$ number. For large
$\tan\beta$ this coefficient will remain as $O(10)$.

Let us now compare the anomalous $U(1)$ induced EDM with the
right--handed neutrino induced effects pointed out in Ref.
\cite{Ellis3} and studied further in Ref. \cite{Masina, peskin}.
The latter effects induce EDM which are given by
\begin{eqnarray}\label{neutrinoEDM1}
d_i\propto [\left(Y^\nu\right)^\dagger \Lambda Y^\nu
,\left(Y^\nu\right)^\dagger Y^\nu],
\end{eqnarray}
where $\Lambda_{ij}=\mbox{log}(M_{GUT}/(M_{\nu^c})_i)\delta_{ij}$.
Here $(M_{\nu^c})_i$ is the mass of the right--handed neutrino of
flavor $i$. With our texture for the neutrino mass matrices
dictated by $U(1)_A$ symmetry we find the right--handed neutrino
induced EDM to be $d_e\sim10^{-29}~e$ cm, which is two to three
orders magnitude smaller than the anomalous $U(1)_A$ induced
effects. In our numerical analysis we present separately our
results for the electron EDM arising from the right--handed
neutrino effects.

\subsection{Numerical Results}

\begin{figure}[!ht]
\vspace*{0.truecm}
\begin{center}
\epsfig{file=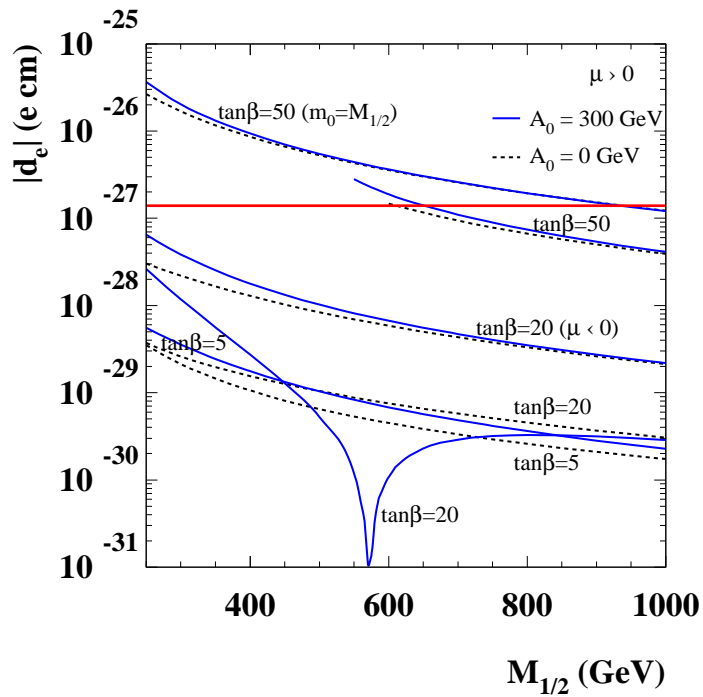,height=3.6in,width=3.6in}
\caption{\footnotesize Electric Dipole Moment of the electron
induced by the flavor gaugino/gauge boson. The (red) horizontal
line shows the current experimental limit on $d_e$. We have chosen
here $m_0=M_{1/2}/4.4$. For $\tan\beta=50$ we show an additional
case with $m_0=M_{1/2}$ (the uppermost curve). For $\tan\beta=20$
and $A_0=300$ GeV we find a cancelation between the $A$--term
contributions given in Eq. (\ref{beta2}) and the soft left/right
mass contributions in Eq. (\ref{betamf}) for our particular fit of
the Yukawa couplings. This cancelation disappears for the choice
of negative $\mu$--term (the curve labeled by
$\mu<0$).}\label{fig1}
\end{center}
\end{figure}

In this sub-section we present our numerical results for the
electron, muon, neutron and the deuteron electric dipole moments.
We adopt the minimal supergravity scenario for supersymmetry
breaking. At the string scale, taken to be $M_{st}=10^{17}$ GeV,
we assume a universal scalar mass $m_0$ and a common gaugino soft
mass $M_{1/2}$. All SUSY breaking parameters
($m_0,\,M_{1/2},\,A_0,\,B_0$) and the $\mu$ term are taken to be
real at the string scale. We choose $\mu>0$ for all cases except
in Fig. \ref{fig1} where we also show results for $\mu <0$. The
anomalous $U(1)$ gauge coupling $g_F$ is chosen to be
$g_{F}^2/4\pi=1/24$, consistent with string unification. The soft
SUSY breaking parameters are evolved from $M_{st}$ to the $U(1)_A$
breaking scale $M_F\simeq M_{st}/50$ (see Eq. (\ref{MAM1}))
including the $U(1)_A$ gaugino/gauge boson corrections.

We present our results for the EDM for three values of the
parameter $\tan\beta$, small ($5$), medium $(20)$ and large
$(50)$. We take $m_0=M_{1/2}/4.4$ for low and medium values of
$\tan\beta$. This is motivated by the requirement that the right
abundance of cosmological cold dark matter be generated.  With
$m_0=M_{1/2}/4.4$, the right--handed charged sleptons will have
masses slightly above that of the neutralino LSP. The relic
abundance of neutralinos is in the right range with such a
spectrum, as a result of coannihilation \cite{coann}. For large
$\tan\beta$ we also allow the choice $m_0=M_{1/2}$, since
alternative mechanisms for reproducing the right relic abundance
of LSP become available in this case \cite{baer}.

\begin{figure}[!ht]
\vspace*{-0.8truecm}
\begin{center}
\epsfig{file=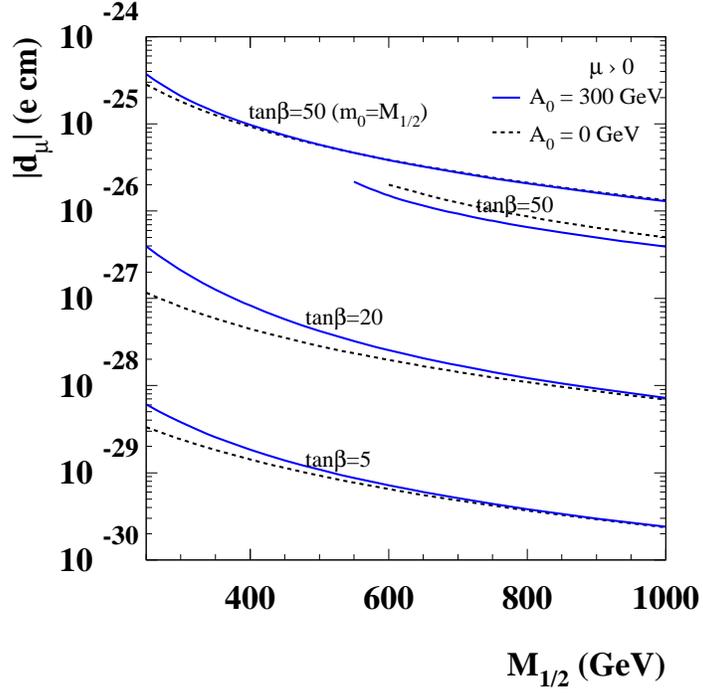,height=3.6in,width=3.6in}
\caption{\footnotesize Electric Dipole Moment of the muon induced
by the flavor gauge corrections. Here $m_0=M_{1/2}/4.4$. For
$\tan\beta = 50$, we also present results for the case
$m_0=M_{1/2}$.}\label{fig2}
\end{center}
\end{figure}

\begin{figure}[!ht]
\begin{center}
\epsfig{file=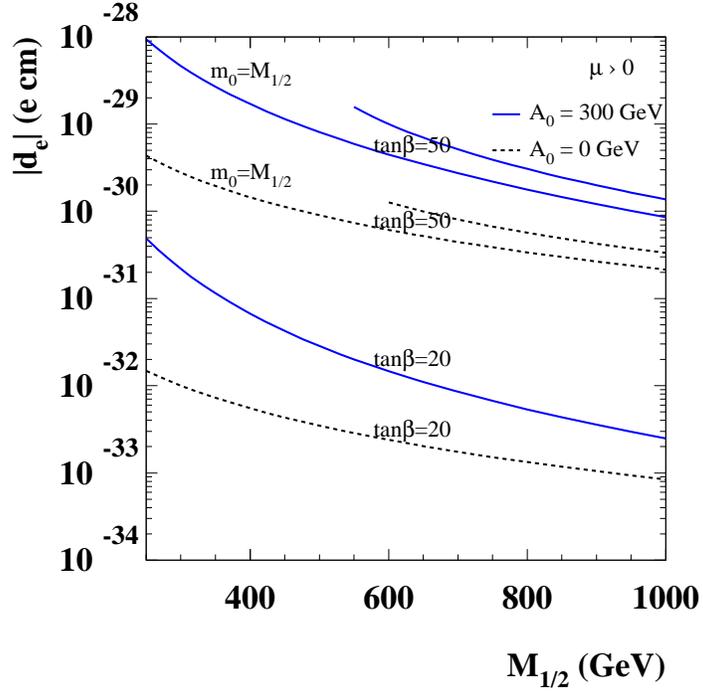,height=3.6in,width=3.6in}
\caption{\footnotesize Electric Dipole Moment of the electron
induced purely by the right--handed neutrino threshold
corrections. The notation is the same as in Fig. \ref{fig1}.
}\label{fig3}
\end{center}
\end{figure}

\begin{figure}[!ht]
\vspace*{-.5truecm}
\begin{center}
\epsfig{file=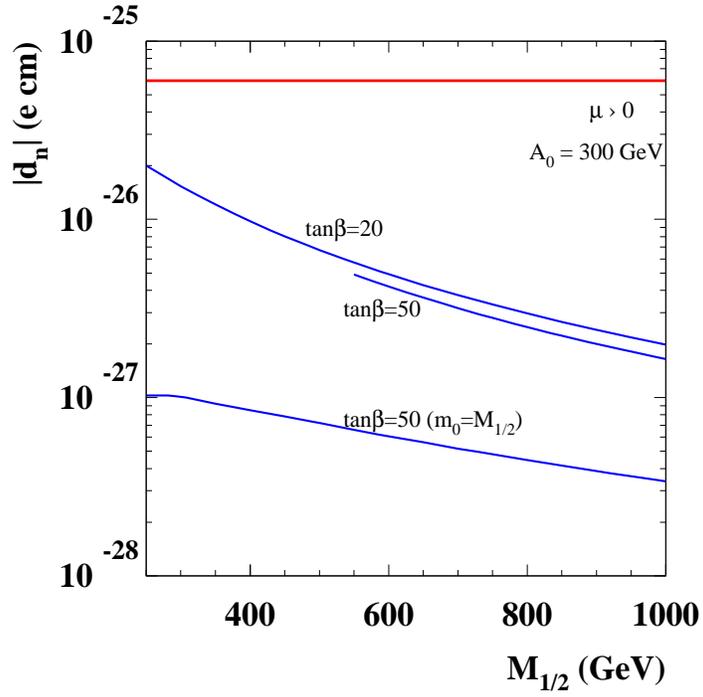,height=3.6in,width=3.6in}
\caption{\footnotesize Electric Dipole Moment of the neutron
induced by the flavor gaugino/gauge boson corrections. Here
$m_0=M_{1/2}/4.4$, with an additional case $m_0=M_{1/2}$ shown for
$\tan\beta=50$. The horizontal line is the current experimental
limit.}\label{fig4}
\end{center}
\end{figure}

\begin{figure}[!ht]
\vspace*{-.5truecm}
\begin{center}
\epsfig{file=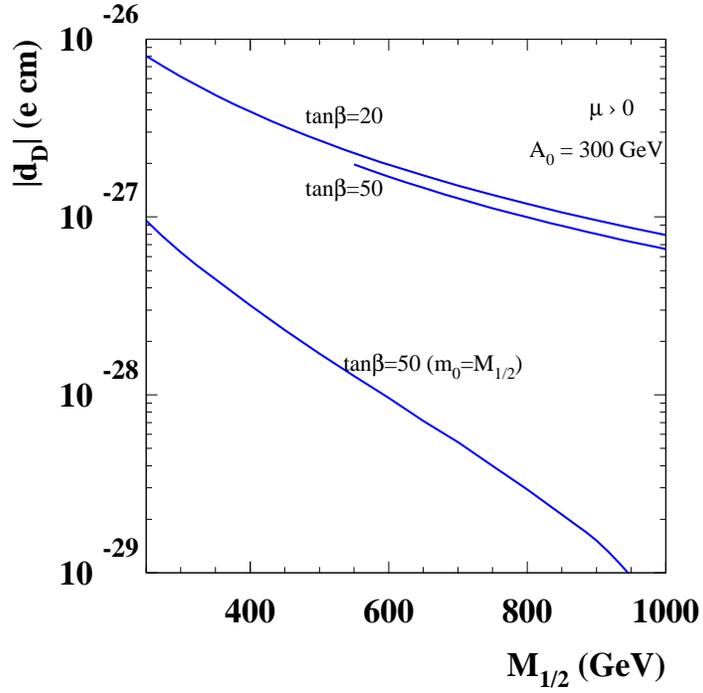,height=3.6in,width=3.6in}
\caption{\footnotesize Elelctric Dipole Moment of the deuteron
induced by the flavor gaugino/gauge boson corrections. Here
$m_0=M_{1/2}/4.4$, with an additional case $m_0=M_{1/2}$ shown for
$\tan\beta=50$.}\label{fig5}
\end{center}
\end{figure}

We vary $M_{1/2}$ in the range $250$ GeV to $1$ TeV. The results
are presented for two different values of $A_0$, $0$ and $300$
GeV. The lepton EDMs induced by the flavor $U(1)$ gaugino/gauge
boson contribution are plotted against the universal gaugino mass
$M_{1/2}$ in Figures \ref{fig1}-\ref{fig2}. In Figure \ref{fig3}
the electron EDM induced  purely  by the right--handed neutrino
effects is plotted. In Figure \ref{fig4}-\ref{fig5} we plot the
EDM of the neutron and the deuteron arising from the flavor $U(1)$
gauge boson/gaugino effects.

As input at $M_{st}$ we choose the Yukawa coupling matrices given
in Eqs. (\ref{Yukup})--(\ref{Yukneutrino}) of Appendix \ref{a2}
(for $\tan\beta=5$). These are obtained by extrapolating the low
energy Yukawa couplings to $M_{st}$ and applying bi-unitary
transformations  at $M_{st}$ to generate the texture given in Eq.
(\ref{massM1}). The low energy Yukawa couplings and their
extrapolation are discussed in Appendix \ref{a2}. As for the
neutrino Dirac Yukawa couplings, we choose $Y^\nu$ to be such that
in the flavor basis (after the bi--unitary rotations) it exhibits
approximately the structure given in Eq. (\ref{massM1}) with
$(Y^\nu)_{33}\sim\e^p$. For a given choice of hierarchical
light--neutrino spectrum this would uniquely fix the right--handed
neutrino mass matrix through the seesaw mechanism. $M_{\nu^c}$
will then have the form given in Eq. (\ref{massM1}). We set
$(M_{\nu^c})_{33}=M_R^0 \epsilon^{2p}$ with $M_R^0 \simeq 4 \times
10^{14}$ GeV. The eigenvalues of the right--handed neutrino mass
matrix are important for the lepton EDMs induced by the
right--handed neutrino threshold effects. It should be noted that
the unitary rotations applied on the diagonal Yukawa matrices at
$M_{st}$ are not unique, except that they should conform to the
fermion mass matrix structure shown in Eq. (\ref{massM1}). So our
fits should be taken only as indicative, and not definitive. We
expect differences of order one in our numerical results on EDM
arising from the arbitrariness in these unitary matrices.

In Figure \ref{fig1} the electron EDM induced by the $U(1)_A$
gaugino/gauge boson contributions to the soft masses and
$A$--terms are plotted as a function of $M_{1/2}$ for three values
of $\tan\beta$. We see that some parts of the parameter space are
already excluded by the current experimental upper bound $d_e\leq
1.6\times10^{-27}~e$ cm and that the other parts are in the range
which will be tested by next generation electron EDM experiments
\cite{DeMille}.

In Figure \ref{fig2} we plot the muon EDM as a function of SUSY
breaking parameters. We find $d_\mu$ to be in the range
$(10^{-25}-10^{-28})~e$ cm for most of the parameter space. This
value is somewhat smaller than than $d_e(m_\mu/m_e)$, which would
be the naive expectation based on the scaling of lepton masses.
This happens for the following reason. The second and third family
left--handed charged sleptons have the same $U(1)_A$ charge, so
the flavor gauge bosons/gaugino will  not generate any mass
splitting between these sleptons. The mixing in the right--handed
charged slepton sector is suppressed by a factor $\e^2$ for all
$\tan\beta$, compared to the suppression factor $\e$ between the
first and the second generations. On the other hand, we find quite
an enhancement of the muon EDM for the choice $m_0=M_{1/2}$ and
$\tan\beta=50$.  For this choice,  the electron EDM is well above
the experimental bound. Since the two EDMs are induced by
independent phases, it is possible to choose the parameters such
that the electron EDM is below the experimental limit and at the
same time the muon EDM is at the level of
$\sim(10^{-25}-10^{-24})\, e$ cm, although we do not attempt such
an explicit solution here. It should also be pointed out that
parts of the parameter space where $d_\mu$ is large is already
ruled out by the experimental upper limit for the radiative decay
$\mu\rightarrow e\gamma$ for the numerical fits shown \cite{beg}.
The remaining regions will be put to experimental scrutiny by
future experiments \cite{MuonEDM}.

In Figure \ref{fig3} we present for comparison, the electron EDM
arising solely from the right--handed neutrino threshold effects
\cite{Ellis3}.  With the proper decoupling of the right--handed
neutrinos \cite{peskin} we find our results to be in rough
agreement with those in Ref. \cite{Ellis3, Masina, peskin}.
Nevertheless these effects, which yield at most $d_e \sim
10^{-29}e\,\mbox{cm}$,  are much smaller compared to the $U(1)_A$
effects.

In Figure \ref{fig4} we plot the neutron EDM versus $M_{1/2}$. In
Figure \ref{fig5} we plot the deuteron EDM.  Details of the
calculations are given in Appendix \ref{a3}. In both cases our
numerical results are in the interesting range which should be
accessible to proposed experiments in the near future. We find the
contributions from the CKM phase to be of the same order as the
contributions from the $U(1)_A$ gaugino/gauge boson sector.
Figures \ref{fig4}--\ref{fig5} include both these effects.  The
flavor sector contribution to the  neutron EDM is somewhat smaller
compared to the leptonic EDM due to the gluino focusing effect.
(The squarks receive flavor universal contributions for their
masses below $M_F$ from the gluino, which tends to suppress flavor
violation and thus $d_n$.)

We have also studied the constraint on the chromoelectric dipole
moment for the strange quark $d^C_s$ arising from $^{199}$Hg EDM
\cite{Fortson, HisanoShimuzu}. This bound reads as $|d^C_s|\leq
5.8\times10^{-25}~e\,\mbox{cm}.$ This constraint is easily
satisfied in our model. The down--type squark mixing in the $(23)$
sector is suppressed by a factor $\e^2$ for the right--handed
squarks, and is vanishing to leading order for the left--handed
squarks, similar to the case of $\mu-\tau$ mixing. Consequently,
we find the chromoelectric EDM of the strange quark to be about
two to three orders of magnitude below the experimental limit.

The soft SUSY breaking bilinear $B$--term and the gaugino masses
will develop complex phases via the one--loop and two--loop RGE
corrections respectively arising from the $A$--term contributions.
In our model we find these corrections to be negligible compared
to the $U(1)_A$ flavor gaugino/gauge boson effects.

\newpage
\newpage
\newpage
\section{Conclusions}
In this paper we have studied the electric dipole moments of the
electron, muon and the neutron induced by a flavor dependent
$U(1)$ symmetry which explains the hierarchy of fermion masses and
mixings in a natural way via the Froggatt-Nielsen mechanism.  This
$U(1)$ symmetry may be identified as the anomalous $U(1)$ of
string theory.  This symmetry is broken spontaneously at a scale
$M_F$ slightly below the string scale, $M_F \sim M_{st}/50$. In
the momentum regime $M_F \leq \mu \leq M_{st}$, the flavor
$U(1)_A$ gauge boson sector will be active and will contribute to
the soft SUSY breaking parameters in a flavor dependent fashion.
We adopt the minimal supergravity scenario for SUSY breaking, and
assume that the soft SUSY breaking parameters are universal and
real.  The complex Yukawa couplings will still induce phases in
the soft SUSY masses and the $A$-parameters, leading to the
generation of EDM.  This is the main source of the EDM that we
have studied here.

We have presented our numerical results for the electron, muon,
neutron and the deuteron EDMs in Figures \ref{fig1}--\ref{fig5} as
functions of supersymmetry breaking parameters.  $d_e$ and $d_n$
are very close to the current experimental limits, $d_e \sim
(10^{-26}-10^{-27}) ~e$ cm and $d_n \sim 10^{-27}~e$ cm.  For the
deuteron, our prediction, $d_D \sim (10^{-26}-10^{-27})~e$ cm,
would make it within reach of proposed experiments. For the case
of the muon, although $d_\mu$ is rather small for low $\tan\beta$,
in the case of large $\tan\beta \sim 50$, for certain choices of
phases in the Yuakawa couplings, we have found the induced  EDM to
be as large as $d_\mu \sim (10^{-25}- 10^{-24})~e$ cm, which might
be accessible to future experiments \cite{MuonEDM}.  In the
leptonic sector, these EDMs are much larger than the ones induced
by the neutrino seesaw sector, which yields, for example, $d_e
\sim 3 \times 10^{-29}~e$ cm with our texture of fermion mass
matrices dictated by flavor symmetries. In Figure \ref{fig4} we
present our results for the induced $d_e$ arising from the
neutrino seesaw sector. Discovery of electric dipole moments for
the electron, muon and the neutron can shed light on one of the
fundamental puzzles of Nature, viz., the origin of mass for
elementary particles.

\section*{Acknowledgments}

We thank professor Lee Roberts and the Physics Department of
Boston University for hospitality during our participation in the
International Symposium on ``Lepton Moments", Cape Cod, MA, June
2003. Ts. E. also acknowledges financial support for
participation. This work is supported in part by DOE Grant \#
DE-FG02-04ER46140, and an award from the Research Corporation.

\begin{appendix}
\section{Appendix}

\subsection{RGE for Soft Parameters including $U(1)_A$
Corrections}\label{a1}

Here we list all the $\beta$--functions for the soft masses and
$A$--terms including contributions from the anomalous $U(1)$
gaugino/gauge boson sector (see Ref. \cite{beg} for details).

The one--loop $\beta$--functions for the soft sfermion masses are
given by
\begin{eqnarray}\label{betamL}
\beta\left(m_{\tilde{f}}^2\right)_{ij}=\beta\left(m_{\tilde{f}}^2\right)^{MSSM}_{ij}
+\beta\left(m_{\tilde{f}}^2\right)^{A}_{ij}+\beta\left(m_{\tilde{f}}^2\right)^{N}_{ij},
\end{eqnarray}
where the superscripts $MSSM$, $A$ and $N$ denote the
contributions from the MSSM sector, the anomalous $U(1)$ gauge
sector, and the right--handed neutrino sector respectively.
$\beta\left(m_{\tilde{f}}^2\right)^{N}_{ij}$ are present only for
the left--handed slepton and the sneutrino soft masses and their
explicit form is given elsewhere. The anomalous $U(1)$ part
$\beta\left(m_{\tilde{f}}^2\right)^{A}_{ij}$ is \cite{martin1}
\begin{eqnarray}\label{betamf}
\beta\left(m_{\tilde{f}}^2\right)^A_{ij}=\frac{1}{16\pi^2}
2q^f_ig_F^2\delta_{ij}\left(\sigma-4q^f_i(M_{\lambda_F})^2\right).
\end{eqnarray}
Here $\sigma$ is defined as
\begin{eqnarray}\label{sigma}
&&\sigma=3\,{\rm
Tr}\left(2q^Q{\tilde{m}_Q}^2+q^u{\tilde{m}_u}^2+q^d{\tilde{m}_d}^2\right)
+{\rm
Tr}\left(2q^L{\tilde{m}_L}^2+q^e{\tilde{m}_{e}}^2+q^\nu{\tilde{m}_\nu}^2\right)\nn\\&&\hspace{2cm}+q_s\tilde{m}_s^2+\sum_k
q^X_k{\tilde{m}_{X_k }}^2\, ,
\end{eqnarray}
where $\tilde{m}_{X_k}$ are the soft masses of the extra particles
$X_k$ introduced for  anomaly cancelation via the Green--Schwarz
mechanism. The trace is taken over family space. Here
$\beta\left(\tilde{m}^2_L\right)^{MSSM}_{ij}$ etc., stand for the
MSSM $\beta$--functions without the neutrino and the flavor
$U(1)_A$ contributions.

Introducing a notation
\begin{eqnarray}\label{NotaA}
A^f_{ij}\equiv a^f_{ij}\e^{n^f_{ij}},
\end{eqnarray}
the $U(1)$ gaugino part of $A$--term $\beta$--functions is given
by
\begin{eqnarray}\label{beta2}
\beta(A^f)^A_{ij}=-\frac{1}{8\pi^2}g_F^2A^f_{ij}\left((q^{f}_i)^2+(q^{f^c}_j)^2+h^2\right)
+\frac{1}{4\pi^2}g_F^2\,Z^f_{ij}Y^f_{ij}M_{\lambda_F}\,,
\end{eqnarray}
where $g_F$ and $M_{\lambda_F}$ are the $U(1)_A$ gauge coupling
and the gaugino mass respectively. $h$ is the $U(1)_A$ charge of
the up--type (down--type) higgs doublet if $f^c$ is up--type
(down--type).  Here we defined the combination of the $U(1)_A$
charges $Z^f_{ij}$ as
\begin{eqnarray}\label{chargec1}
Z^f_{ij}&=&q^{f}_i q^{f^c}_j+q^{f}_i h +q^{f^c}_j h
+n^f_{ij}q_s(q^{f}_i+q^{f^c}_j+
h)+\frac{1}{2}n^f_{ij}(n^f_{ij}-1)q_s^2.
\end{eqnarray}
From the $U(1)_A$ charge assignments for the MSSM fields as given
in Table \ref{tcharge} one has
\begin{eqnarray}\label{Ze1}
&&Z^e=-\pmatrix{(\,11,\,13,\,16)\,&(\,4,\,6,\,9)&(\,1,\,3,\,6)\cr
(\,10,\,11,\,13)&(\,3,\,4,\,6)&(\,0,\,1,\,3)\cr
(\,10,\,11,\,13)&(\,3,\,4,\,6)&(\,0,\,1,\,3)}\ ,
\end{eqnarray}
\begin{eqnarray}\label{Zd1}
&&Z^d=-\pmatrix{(\,11,\,13,\,16)\,&(\,9,\,10,\,12)&(\,9,\,10,\,12)\cr
(\,4,\,6,\,9)&(\,3,\,4,\,6)&(\,3,\,4,\,6)\cr
(\,1,\,3,\,6)&(\,0,\,1,\,3)&(\,0,\,1,\,3)}\ ,
\end{eqnarray}
\begin{eqnarray}\label{Zn1}
&&Z^\nu=-\pmatrix{(\,2,\,4,\,7)\,&(\,1,\,3,\,6)&(\,1,\,3,\,6)\cr
(\,1,\,2,\,4)&(\,0,\,1,\,3)&(\,0,\,1,\,3)\cr
(\,1,\,3,\,4)&(\,0,\,1,\,3)&(\,0,\,1,\,3)}\ ,
\end{eqnarray}
for the three different values of $p=(0,\,1,\,2)$ and
\begin{eqnarray}\label{Zu1}
Z^u=-\pmatrix{20&13&10\cr 13&6&3\cr 10&3&0}\ ,\hspace{.5cm}
\end{eqnarray}
independent of $p$.

The terms we are interested in are the ones proportional to the
$U(1)_A$ gaugino mass $M_{\lambda_F}$  and the term proportional
to $\sigma$ in the $\beta$--functions. Beside these, the $A$--term
$\beta$--functions contain a flavor dependent piece which arises
from the wave--function renormalization. Since the corresponding
Yukawa $\beta$--functions contain the same terms, these are
simultaneously diagonalized, and do not lead to flavor violation.

\subsection{Fermion Mass Fit}\label{a2}

Here we present the numerical fits to the fermion masses and
mixings adopted for the calculation of the EDMs.  As input at low
energy we choose the following values for the running quark masses
\cite{Gasser}:
\begin{eqnarray}\label{QuarkM}
&&m_u(\mbox{1\,GeV})=5.1\,\mbox{MeV},\hspace{.5cm}
m_c(m_c)=1.27\,\mbox{GeV},\hspace{.5cm} m_t(m_t)=167\,\mbox{GeV},\nn\\
&&m_d(\mbox{1\,GeV})=8.9\,\mbox{MeV},\hspace{.5cm}
m_s(\mbox{1\,GeV})=130\,\mbox{MeV},\hspace{.5cm}
m_b(m_b)=4.25\,\mbox{GeV}.
\end{eqnarray}
The CKM mixing matrix is chosen in the standard parametrization
with $\theta_{12}=0.221$, $\theta_{13}=0.005$, $\theta_{23}=0.043$
and the complex phase $\delta=0.86$.  We Use two--loop QED and QCD
renormalization group equations to evolve these masses from the
low energy scale to the scale of SUSY breaking. We obtain the
following running factor $r_f\equiv m_f(M_{SUSY})/m_f(m_f)$ for
the fermion masses at the SUSY breaking scale $M_{SUSY}$,
initially chosen to be $500$ GeV with $\alpha_s(M_Z)=0.118$:
\begin{eqnarray}\label{RunningF}
(r_t,\,r_b,\,r_\tau,\,r_u,\,r_c,\,r_{d,s},\,r_{e,\mu})=(0.943,0.605,0.991,0.395,0.442,0.395,0.398).
\end{eqnarray}
Then these masses at $M_{SUSY}$ are used to calculate the Yukawa
couplings in $\overline{DR}$ scheme. Using one--loop SUSY RGE
evolution above $M_{SUSY}$ we obtain the Yukawa couplings at the
$U(1)_A$ breaking scale ($M_F \sim10^{15}$ GeV) to be
\begin{eqnarray}\label{YukawaHigh1}
&&(Y_u,\,Y_c,\,Y_t)=(5.2699\times10^{-6},\,1.4634\times10^{-3},\,0.55498),\nn\\
&&(Y_d,\,Y_s,\,Y_b)=(3.4415\times10^{-5},\,5.0222\times10^{-4},\,2.8247\times10^{-2}),\nn\\
&&(Y_e,\,Y_\mu,\,Y_\tau)=(1.0388\times10^{-5},\,2.1481\times10^{-3},\,3.6239\times10^{-2}),\nn\\
&&(Y_{\nu_1},\,Y_{\nu_2},\,Y_{\nu_3})=(7.4112\times10^{-4},\,3.5135\times10^{-3},\,4.7212\times10^{-2}),
\end{eqnarray}
for $\tan\beta=5$,
\begin{eqnarray}\label{YukawaHigh2}
&&(Y_u,\,Y_c,\,Y_t)=(5.0919\times10^{-6},\,1.4140\times10^{-3},\,0.53090),\nn\\
&&(Y_d,\,Y_s,\,Y_b)=(1.3834\times10^{-4},\,2.0189\times10^{-3},\,0.11508),\nn\\
&&(Y_e,\,Y_\mu,\,Y_\tau)=(4.1778\times10^{-5},\,8.6456\times10^{-3},\,0.14818),\nn\\
&&(Y_{\nu_1},\,Y_{\nu_2},\,Y_{\nu_3})=(3.5720\times10^{-3},\,1.6934\times10^{-2},\,0.2275),
\end{eqnarray}
for $\tan\beta=20$ and
\begin{eqnarray}\label{YukawaHigh3}
&&(Y_u,\,Y_c,\,Y_t)=(5.3161\times10^{-6},\,1.4764\times10^{-3},\,0.59610),\nn\\
&&(Y_d,\,Y_s,\,Y_b)=(4.2226\times10^{-4},\,6.1621\times10^{-3},\,0.41186),\nn\\
&&(Y_e,\,Y_\mu,\,Y_\tau)=(1.2720\times10^{-4},\,2.6645\times10^{-2},\,0.51807),\nn\\
&&(Y_{\nu_1},\,Y_{\nu_2},\,Y_{\nu_3})=(1.7822\times10^{-2},\,8.4492\times10^{-2},\,1.1354),
\end{eqnarray}
for $\tan\beta=50$.

We have chosen the Dirac neutrino Yukawa couplings such that in
the flavor basis (after the bi--unitary rotations) it exhibits the
assumed structure in Eq. (\ref{massM1}). The light--neutrino
masses and the lepton mixing matrix are chosen to be
$m_{\nu_1}=2.7\times10^{-3}$ eV, $m_{\nu_2}=6.4\times10^{-3}$ eV
and $m_{\nu_3}=8.6\times10^{-2}$ eV and
\begin{eqnarray}\label{VMNS}
&&V_{MNS}=\pmatrix{0.8494&-0.5262&-0.04\cr
0.3915&0.5775&0.7164\cr-0.3539&-0.6242&0.6965}\,,
\end{eqnarray}
which are compatible with the observed light--neutrino oscillation
parameters. From this the right--handed neutrino mass matrix is
fixed uniquely with $(M_{\nu^c})_{33}\simeq M^0_R\e^{2p}$ where
$M^0_R=4\times10^{14}$ GeV. The eigenvalues of $M_{\nu^c}$ are
used to calculate the lepton EDM induced by the right--handed
neutrino threshold effects.

 In the following, we present our fits to the texture of Eq. (\ref{massM1}) which
 have been used in our numerical calculations for $\tan\beta=5$
 (We have similar fits upto an overall factor for $\tan\beta=20,\,50$).
 This fit is not unique, and so  can lead to order one uncertainty
 in our EDM results. The following fit is found by applying
 bi-unitary transformations with complex phases on the
 diagonal Yukawa coupling matrices at the $U(1)_A$ breaking scale $M_F$.
 We introduce a notation for the Yukawa couplings:
\begin{eqnarray}\label{NotaYuk}
Y^f_{ij}\equiv y^f_{ij}\e^{n^f_{ij}}.
\end{eqnarray}
At $M_F$ we find the following fit (for $\tan\beta = 5)$:
\begin{eqnarray}\label{Yukup}
Y^u=\pmatrix{(1.45+1.60\,i)\,\e^8&(-0.563-1.24\,i)\,\e^6&(1.50-0.397\,i\,)\,\e^4\cr
(-0.769-0.584\,i)\,\e^6&(0.765-0.109\,i)\,\e^4&(-0.255-0.261\times10^{-2}\,i)\,\e^2\cr
(-0.282-0.204\,i)\,\e^4&(0.274-4.40\times10^{-2}\,i)\,\e^2&0.554-2.80\times10^{-5}\,i},\hspace{.5cm}
\end{eqnarray}
\begin{eqnarray}\label{Yukdown}
Y^d=\e^2\pmatrix{(1.87-1.69\,i)\,\e^5&(1.93+0.849\,i)\,\e^4&(1.29+0.957\,i)\,\e^4\cr
(-0.404-0.248\,i)\,\e^3&(0.552+1.54\times10^{-2}\,i)\,\e^2&(0.702-0.546\,i)\,\e^2\cr
(-1.52-0.435\,i)\,\e&0.312+0.314\,i&0.543-4.74\times10^{-4}\,i},\hspace{.5cm}
\end{eqnarray}
\begin{eqnarray}\label{Yuklep}
Y^e=\e^2\pmatrix{(3.52\times10^{-2}+0.480\,i)\e^5&(-1.85-1.74\,i)\,\e^3&(-0.539-0.579\,i)\,\e\cr
(-0.170-0.612\,i)\e^4&(1.15-4.65\times10^{-2}\,i)\,\e^2&0.319-0.321\,i\cr
(0.538-0.421\,i)\e^4&(-0.419-0.536\,i)\,\e^2&0.784-9.73\times10^{-4}\,i},\hspace{.5cm}
\end{eqnarray}
\begin{eqnarray}\label{Yukneutrino}
Y^\nu=\e^2\pmatrix{(0.232-0.190\,i)\,\e^2&(0.217-6.09\times10^{-2}\,i)\,\e&(-0.206-0.637\,i)\,\e\cr
(0.638-0.652\,i)\,\e&-7.82\times10^{-2}+0.537\,i&0.804+0.296\,i\cr
(0.305-0.392\,i)\,\e&-4.41\times10^{-3}+0.277\,i&0.404-3.89\times10^{-2}\,i}.
\end{eqnarray}
Note that all coefficients multiplying $\e^{n_{ij}}$ in Eqs.
(\ref{Yukup})--(\ref{Yukneutrino}) are of order unity.

\subsection{Formulas for Electric Dipole Moments}\label{a3}

We list here the formulas for the electric dipole moments of
leptons and quarks in the MSSM from Ref. \cite{Ibrahim1}, which we
have used in our numerical analysis.

The EDMs of elementary fermions are sum of neutralino, chargino
and for quarks gluino contributions which we denote as $d^N_f$,
$d^C_f$ and $d^G_q$. In addition to these, the  quarks receive
contributions from chromoelectric  and purely gluonic
dimension--six operators \cite{weinberg}. We have not considered
here the latter one, since these effects turn out to be small. The
effective EDM operator $d_f$ for a spin--$\frac{1}{2}$ particle is
given by
\begin{eqnarray}\label{LagEDM}
{\cal L}=-\frac{i}{2}d_f{\bar\psi}\sigma_{\mu\nu}\gamma_5\psi
F^{\mu\nu}
\end{eqnarray}
The EDM $d_f$ in general has the following components in a
supersymmetric theory:
\begin{eqnarray}\label{FormulEDM}
d^N_{f_i}/e&=&\frac{\alpha}{8\pi\sin^2\theta_W}\sum^6_{x=1}\sum^4_{a=1}
\mbox{Im}\left(N^{f_i}_{xa}\right)\frac{M_{\chi^0_a}}{m^2_{\tilde{f_x}}}Q_{\tilde{f}_x}
A\left(\frac{M^2_{\chi^0_a}}{m^2_{\tilde{f_x}}}\right),\nn\\
d^C_{u}/e&=&\frac{-\alpha}{8\pi\sin^2\theta_W}\sum^6_{x=1}\sum^2_{b=1}
\mbox{Im}\left(C^{u}_{xb}\right)\frac{M_{\tilde{\chi}^+_b}}{m^2_{\tilde{d}_x}}
\left(B\left(\frac{M_{\tilde{\chi}^+_b}^2}{m^2_{\tilde{d}_x}}\right)
-\frac{1}{3}A\left(\frac{M_{\tilde{\chi}^+_b}^2}{m^2_{\tilde{d}_x}}\right)\right),\nn\\
d^C_{d}/e&=&\frac{-\alpha}{8\pi\sin^2\theta_W}\sum^6_{x=1}\sum^2_{b=1}
\mbox{Im}\left(C^{d}_{xb}\right)\frac{M_{\tilde{\chi}^+_b}}{m^2_{\tilde{u}_x}}
\left(\frac{2}{3}A\left(\frac{M_{\tilde{\chi}^+_b}^2}{m^2_{\tilde{u}_x}}\right)
-B\left(\frac{M_{\tilde{\chi}^+_b}^2}{m^2_{\tilde{u}_x}}\right)\right),\nn\\
d^G_{q_i}/e&=&-\frac{\alpha_s}{3\pi}\sum^6_{x=1}
\mbox{Im}\left(G^{q_i}_{x}\right)\frac{M_{\tilde{g}}}{m^2_{\tilde{q}_x}}Q_{\tilde{q}_i}
A\left(\frac{M_{\tilde{\chi}^+_b}^2}{m^2_{\tilde{u}_x}}\right),
\end{eqnarray}
where
\begin{eqnarray}\label{FuncEDM}
A(X)&=&\frac{1-X^2+2X\log X}{\left(1-X\right)^3},\nn\\
B(X)&=&\frac{3-4X+X^2+2\log X}{\left(1-X\right)^3}.
\end{eqnarray}
Here $M_{\tilde{\chi}^0_a}$, $M_{\tilde{\chi}^+_b}$ and
$M_{\tilde{g}}$ are the neutralino, chargino and the gluino masses
respectively. $m^2_{\tilde{f}_{x}}$ ($x=1,...,6$) are the
eigenvalues of the sfermion mass matrices. The coefficients
$N^f_{xa}$, $C^f_{xb}$ and $G^q_x$ are given by
\begin{eqnarray}\label{coeffEDM}
N^{f_i}_{xa}&=&\left[\sqrt{2}\tan\theta_WQ_{f_i}
\left(O^N\right)_{1a}U^f_{i+3,x}
-K_f\left(O^N\right)_{a^\prime a}U^f_{i,x}\right]\nn\\
&\times&\left[-\sqrt{2}\left\{\tan\theta_W\left(Q_{f_i}-
T_{f_i}\right)\left(O^N\right)_{a1}+T_{3f_i}\left(O^N\right)_{a2}\right\}
U^{f*}_{i,x}\right.\nn\\&-&\left. K_f\left(O^N\right)_{a^\prime
a}U^{f*}_{i+3,x}\right],\nn\\
C^u_{xb}&=&K_u\left(O^C_R\right)^*_{b2}U^d_{1,x}
\left[\left(O^C_L\right)_{b1}U^{d}_{4,x}-K_d\left(O^C_L\right)_{b
2}U^{d}_{4,x}\right]^*,\nn\\
C^d_{xb}&=&K_d\left(O^C_L\right)^*_{b2}U^u_{i1}
\left[\left(O^C_R\right)_{b1}U^{u}_{1,x}-K_u\left(O^C_R\right)_{b
2}U^{u}_{4,x}\right]^*,\nn\\
G^{q_i}_x&=&U^q_{i,x}U^{q*}_{i+3,x},
\end{eqnarray}
where $K_u=m_u/(\sqrt{2}M_W\sin\beta)$ and
$K_{l,d}=m_{l,d}/(\sqrt{2}M_W\cos\beta)$. $O^N$ and
$O^C_L,\,O^C_R$ matrices diagonalize the neutralino and chargino
mass matrices respectively. The index $a^\prime$ of $O^N$ in the
neutralino contribution formula takes value of $3(4)$ for
$T_{3f}=-\frac{1}{2}(\frac{1}{2})$. The chromoelectric dipole
moments $\tilde{d}_q$ for quarks are defined as
\begin{eqnarray}\label{CEDM}
{\cal L}_{CEDM}=-\frac{i}{2}g_s\tilde{d}_q{\bar
q}T^a\sigma_{\mu\nu}\gamma_5q G^{\mu\nu a}.
\end{eqnarray}
The contributions to $\tilde{d}_q$ from neutralino, chargino and
gluino are given by
\begin{eqnarray}\label{FormulCEDM}
\tilde{d}^N_{q_i}&=&\frac{g^2}{32\pi^2}\sum^6_{x=1}\sum^4_{a=1}
\mbox{Im}\left(N^{q_i}_{xa}\right)\frac{M_{\chi^0_a}}{m^2_{\tilde{q_x}}}
A\left(\frac{M^2_{\chi^0_a}}{m^2_{\tilde{q_x}}}\right),\nn\\
\tilde{d}^C_{q}&=&\frac{-g^2}{32\pi^2}\sum^6_{x=1}\sum^2_{b=1}
\mbox{Im}\left(C^{q}_{xb}\right)\frac{M_{\tilde{\chi}^+_b}}{m^2_{\tilde{q}_x}}
A\left(\frac{M_{\tilde{\chi}^+_b}^2}{m^2_{\tilde{q}_x}}\right),\nn\\
\tilde{d}^G_{q_i}&=&\frac{\alpha_s}{4\pi}\sum^6_{x=1}
\mbox{Im}\left(G^{q_i}_{x}\right)\frac{M_{\tilde{g}}}{m^2_{\tilde{q}_x}}
C\left(\frac{M_{\tilde{\chi}^+_b}^2}{m^2_{\tilde{q}_x}}\right),
\end{eqnarray}
where
\begin{eqnarray}\label{coeffCEDM}
C(X)=\frac{1}{6\left(1-X\right)^2}\left(10X-26+\frac{2X\mbox{log}X}{1-X}-\frac{18\mbox{log}X}{1-X}\right).
\end{eqnarray}
We use the QCD sum rule based estimate of Ref. \cite{pospelov} to
evaluate the neutron and the deuteron EDMs:
\begin{eqnarray}\label{effNDEDM}
&&d_n=0.7\left(d_d-0.25d_u\right)+0.55\left(\tilde{d}_d+0.5\tilde{d}_u\right),\nn\\
&&d_D=0.5\left(d_d+d_u\right)-0.6\left(\tilde{d}_d-\tilde{d}_u+0.3\left(\tilde{d}_d+\tilde{d}_u\right)\right).
\end{eqnarray}
Here the running factors are
$\tilde{d}_q\left(\mbox{1\,GeV}\right)\simeq0.91\tilde{d}_q\left(\mbox{M}_Z\right)$
and $d_q(\mbox{1\,GeV})\simeq 1.2d_q(\mbox{M}_Z)$.
\end{appendix}

\end{document}